\documentclass[letterpaper]{article}

\usepackage[T1]{fontenc}

\usepackage{geometry}
\usepackage{setspace}

\usepackage{achemso}
\setkeys{acs}{articletitle = true, maxauthors=0}
\usepackage{graphicx}
\usepackage{float}
\newfloat{scheme}{htbp}{los}
\floatname{scheme}{Scheme}
\floatname{chart}{Chart}
\newfloat{graph}{htbp}{loh}

\usepackage{chemformula} 
\usepackage[version = 4]{mhchem} 

\usepackage{authblk}
\author[1,2]{Martín Aversa}
\author[1]{Florencia Cammarata}
\author[1,3,4]{Camila Borrazás}
\author[5,6]{Nadia C. Vega}
\author[5,6]{David Comedi}
\author[1,2]{Andrea V. Bragas}
\author[1,2]{Gustavo Grinblat*}

\affil[1]{Universidad de Buenos Aires. Facultad de Ciencias Exactas y Naturales. Departamento de Física. Buenos Aires, 1428, Argentina}
\affil[2]{CONICET-Universidad de Buenos Aires. Instituto de Física de Buenos Aires (IFIBA). Buenos Aires, 1428, Argentina}
\affil[3]{Gerencia Química e INN, CNEA. Av. Gral. Paz 1499, San Martín, Buenos Aires, 1650, Argentina. CONICET}
\affil[4]{Instituto Sabato, UNSAM-CNEA. Av. Gral. Paz 1499, San Martín, Buenos Aires, 1650, Argentina}
\affil[5]{NanoProject - LAFISO y LN, Facultad de Ciencias Exactas y Tecnología, Universidad Nacional de Tucumán, Avenida Independencia 1800, San Miguel de Tucumán, T4002BLR Tucumán, Argentina}
\affil[6]{Instituto de Física del Noroeste Argentino - INFINOA (CONICET-UNT), San Miguel de Tucumán, Tucumán, Argentina}

\title{Localized Strain Tuning of Second-Harmonic Generation in MoS$_2$ Monolayers and Bilayers}
\date{*Email: grinblat@df.uba.ar}

\begin{document}

\maketitle

\begin{abstract}
We investigate the modulation of second-harmonic generation (SHG) in CVD-grown MoS$_2$ monolayers and few-layer flakes through localized mechanical deformation applied by an atomic force microscope (AFM) tip. MoS$_2$ membranes suspended over microstructured Si$_3$N$_4$ substrates allow simultaneous measurements of force and SHG intensity during controlled approach and retraction cycles. Suspended regions exhibit enhanced SHG relative to substrate-supported regions, with pronounced suppression near the clamping boundaries. While the SHG response during compressive loading varies among individual membranes, adhesion-mediated pulling during retraction consistently reduces the SHG intensity until tip detachment, when the signal recovers. By correlating the optical modulation with the measured pull-off force and a phenomenological estimate of effective strain, we obtain a characteristic sensitivity of approximately 12\% SHG modulation per 1\% effective strain, with comparable behavior in monolayers and bilayers. These findings reveal adhesion-mediated mechanical deformation as a distinct and reversible mechanism for controlling nonlinear optical emission in atomically thin membranes.
\end{abstract}

\section*{Keywords}

Transition Metal Dichalcogenides, SHG, Strain Modulation, Van der Waals materials, Nonlinear optics




\vspace{0.5 cm}
The remarkable mechanical flexibility of two-dimensional Transition Metal Dichalcogenides (TMDs), combined with their high fracture thresholds, gives them enormous potential in the field of nanoengineering. The application of mechanical strain ($\varepsilon$) induces changes in their crystal structure that enable the tunable modulation of their optical properties, thus facilitating the design of nanoscale devices for specific applications.

The use of strain in solid-state devices is not exclusive to TMDs; silicon-based technology currently employs strain engineering to optimize charge-carrier mobility in transistors.\cite{Thompson2004} This principle has been successfully extended to TMDs \cite{Datye2022}, demonstrating analogous improvements in electronic transport.

Within the state of the art, applications based on monolayer MoS$_2$ are particularly noteworthy, including photodetectors whose photocurrent response is controlled and enhanced through strain \cite{GANT2019}, as well as visible-range spectrometers with a spectrally tunable response enabled by electromechanical actuation.\cite{Li2025} These advances demonstrate that the gap between fundamental research and the technological implementation of TMDs is becoming increasingly narrow. Nevertheless, there remains considerable room for further investigation into how strain affects other properties, such as magnetic properties \cite{Ren2022}, which could open new avenues in fields such as spintronics.\cite{Mak2012}

In the context of fundamental research, conventional methods consist of applying uniaxial or biaxial strain by bending polymeric substrates.\cite{Carrascoso2021} The application of uniaxial strain using this approach enables the Raman spectrum of MoS$_2$ to be shifted by up to 2.8 cm$^{-1}$ for strain values of up to 1.21\% relative to its equilibrium length.\cite{Yang2014}

The optical property most extensively studied in relation to strain in TMDs is photoluminescence (PL). This effect typically produces a redshift in the spectrum accompanied by a decrease in intensity \cite{He2016,Liu2014}, and transitions from an indirect to a direct bandgap have even been observed under uniaxial strains of 2.5\%.\cite{Wang2015} The study of strain effects localized in specific regions, rather than applied along the entire axis (or axes) of the material, is less prevalent in the literature. However, localized strain has been investigated in the context of PL, yielding consistent results. \cite{Castellanos2013}

The modulation of nonlinear optical effects, such as second-harmonic generation (SHG), has also been extensively studied in TMDs. The high sensitivity of SHG to structural variations, often exceeding that of PL, has also been exploited to assess the structural quality of fabricated samples. \cite{Kourmoulakis2024,xing2024} Strain can induce either an enhancement \cite{Puri2024} or a reduction \cite{Liang2017} in SHG emission intensity, depending on the material. Furthermore, the SHG response depends on the polarization of the excitation source and on the direction of the strain axis. Polarization-resolved SHG responses have been investigated in MoS$_2$, MoSe$_2$, WS$_2$, and WSe$_2$ under uniaxial strains ranging from 0.30\% to 0.48\%.\cite{Mennel2018SHG}

Remarkably, SHG enhancement under biaxial strain has also been demonstrated in monolayer MoS$_2$ through the formation of bubbles.\cite{Liu2023} Likewise, the integration of monolayer MoS$_2$ with titanium dioxide (TiO$_2$) nanowires has been investigated to induce localized strain in heterostructures.\cite{Li2019}

Despite the significant advances in this field, systematic investigations of SHG modulation through localized, non-uniaxial strain are still lacking. In this work, we present results on the modulation of SHG in MoS$_2$ through the controlled application of force with the tip of an atomic force microscope (AFM). These results expand the range of existing techniques for modulating nonlinear optical properties by demonstrating localized control of SHG in suspended MoS$_2$ membranes, with simultaneous monitoring of the optical response and applied mechanical force.

We studied samples of MoS$_2$ flakes transferred onto a substrate containing a 200-nm thick Si$_3$N$_4$ suspended membrane were studied. The substrate features holes of 2~\textmu{}m in diameter that pass entirely through the membrane, enabling the generation of regions of MoS$_2$ decoupled from the substrate. The analyzed samples were grown by chemical vapor deposition (CVD) and transferred onto the substrates using wet (water-assisted)\cite{Borrazas_2025} and dry (viscoelastic stamp-assisted) methods. The thicknesses of these samples were confirmed by measuring their Raman spectra, taking into account that this identification method works independently of the structural phase of the MoS$_2$.\cite{Shi2017} Details on the fabrication, transfer, and characterization of the samples are presented in the Supporting Information section S1.

\begin{figure}
  \centering
  \includegraphics{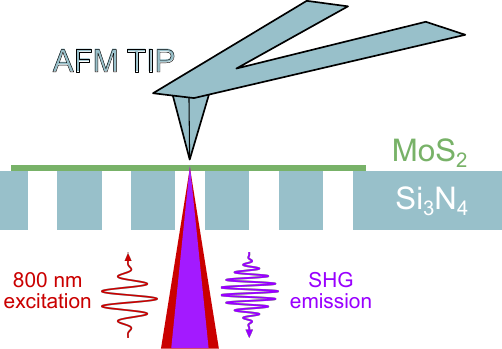}
  \caption{Schematic of the SHG intensity modulation process via controlled and localized strain application. An AFM tip applies pressure in an approach--retraction cycle onto a suspended MoS$_2$ flake on a microstructured Si$_3$N$_4$ substrate. A pulsed Ti:Sapphire laser at 800~nm excites the SHG signal, which is collected in reflection configuration.}
  \label{fig:1}
\end{figure}

For sample analysis, the samples were mounted on a confocal microscope in reflection configuration, using a pulsed Ti:Sapphire excitation laser centered at 800~nm wavelength with a pulse duration of $\sim$100~fs and a repetition rate of 95~MHz. The SHG signal emitted by the material is collected after optical filtering by a PMT detector, allowing SHG maps of the samples to be obtained. The setup is combined at the top with an AFM microscope, which was used to generate controlled deformations via Si$_3$N$_4$ tips on suspended regions of the MoS$_2$ monolayers and bilayers using an approach--retraction cycle of the sample relative to the tip. A schematic of the measurement process is shown in Fig.~\ref{fig:1}. 
To compare the sensitivity of SHG and PL to mechanical deformation, PL maps were also acquired from MoS$_2$ flakes transferred onto microstructured substrates.
No discernible PL-intensity modulation was observed under these conditions, in contrast to SHG (see Section S2 of the Supporting Information for details).

In all cases, the modulation of the obtained signal is quantified as

\begin{equation}
    \textrm{SHG}_\textrm{MOD}=\frac{|\textrm{SHG}_f-\textrm{SHG}_i|}{\textrm{SHG}_i}\times 100,
    \label{ec_shg}
\end{equation}

\noindent where $\textrm{SHG}_f$ and $\textrm{SHG}_i$ represent the SHG intensities measured under the conditions of interest and the corresponding reference conditions, respectively.

\begin{figure}
  \centering
  \includegraphics{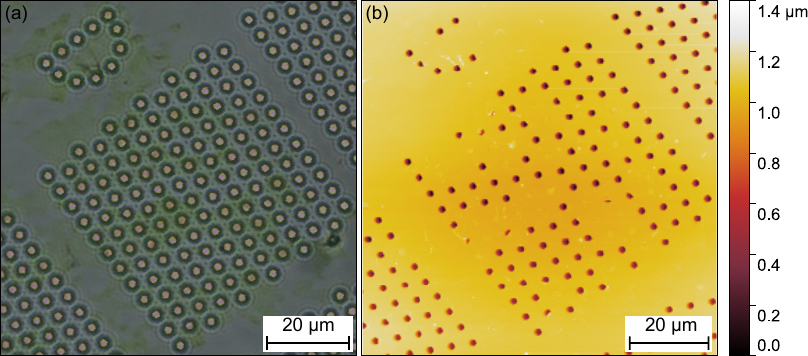}
  \caption{CVD-grown MoS$_2$ flakes transferred onto a microstructured Si$_3$N$_4$ substrate via a polymer-assisted dry transfer method. (a)~Transmission optical microscopy image of the sample, showing poor optical contrast of the flakes over the substrate. (b)~AFM image showing that some holes are covered by suspended MoS$_2$ flakes and therefore do not appear as depressions in the height map.}
  \label{fig:2}
\end{figure}

Fig.~\ref{fig:2}a shows an optical image of a representative sample region, where the strong contrast from the perforations makes it difficult to distinguish the MoS$_2$ monolayers. Nevertheless, the AFM image in Fig.~\ref{fig:2}b confirms that the MoS$_2$ flakes are effectively suspended over the holes, as evidenced by the absence of depth signal in the AFM response over those areas.

\begin{figure}
  \centering
  \includegraphics{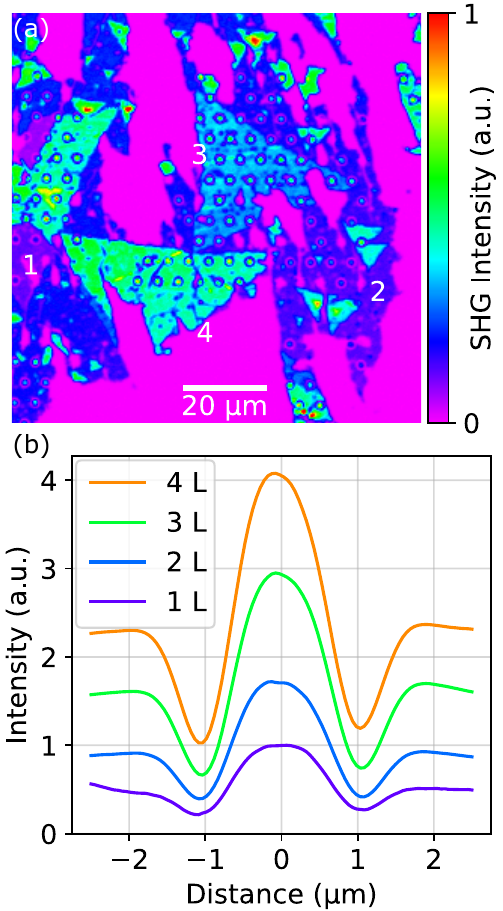}
  \caption{(a)~SHG map of a representative area of CVD-grown MoS$_2$ transferred onto a microstructured Si$_3$N$_4$ substrate. Regions of distinct SHG intensity corresponding to monolayer, bilayer, trilayer, and quadrilayer non-centrosymmetric MoS$_2$ are identified. (b)~Average radial intensity profiles across the suspended regions, grouped by number of layers.}
  \label{fig:3}
\end{figure}

Fig.~\ref{fig:3}a shows an SHG map of MoS$_2$ flakes transferred onto a microstructured Si$_3$N$_4$ substrate using the dry transfer method. Despite some fractures and deformations introduced during transfer, the triangular morphology of the flakes and several suspended regions remain clearly distinguishable. The map reveals pronounced differences in SHG intensity among regions with different numbers of layers, consistent with the influence of both layer thickness and stacking configuration on the nonlinear optical response.\cite{Zhao2016} In 2H-stacked MoS$_2$, inversion symmetry is restored in even-layer structures, strongly suppressing electric-dipole SHG, whereas odd-layer structures remain non-centrosymmetric. In contrast, 3R stacking lacks inversion symmetry regardless of the number of layers, allowing SHG emission from both even- and odd-layer structures.\cite{Shi2017} The strong SHG observed in some multilayer regions is therefore consistent with non-centrosymmetric stacking, possibly 3R. Monolayer, bilayer, trilayer, and quadrilayer regions were identified by combining Raman spectroscopy with SHG intensity maps. Additional examples illustrating SHG responses consistent with different stacking configurations are presented in Section S2 of the Supporting Information.

Beyond the differences associated with layer number and stacking configuration, the SHG map also reveals a pronounced intensity modulation around the circular holes, with a marked reduction near their edges. To quantify this, a circular region-of-interest (ROI) concentric with the holes was defined and 36 radial profiles (taken every 10$^\circ$) were averaged.
Fig.~\ref{fig:3}b shows the average radial SHG intensity profiles obtained from multiple holes for each layer number.
The MoS$_2$ region at the hole edge shows a decrease in SHG emission, while an increase in signal is recorded at the center of the suspended region. This suggests that decoupling the MoS$_2$ from the substrate may enhance SHG emission by relieving pre-existing strain introduced during transfer onto the solid surface. The calculated normal-incidence transmission coefficients for a 200~nm Si$_3$N$_4$ film are 1 and 0.968 at 800~nm and 400~nm, respectively. The corresponding variation in simple optical transmission is therefore only a few percent, substantially smaller than the observed SHG modulation, indicating that transmission losses through the Si$_3$N$_4$ membrane alone cannot account for the measured contrast.

Statistical analysis of the signal modulation at the hole edge relative to the suspended region reveals no significant differences across layer thicknesses. The obtained SHG$_\textrm{MOD}$ values are $(75 \pm 6)\%$ for the monolayer, $(79 \pm 6)\%$ for the bilayer, $(77 \pm 6)\%$ for the trilayer, and $(74 \pm 7)\%$ for the quadrilayer. Furthermore, the deformation of the suspended membrane induced by gravitational loading was estimated to be approximately ten orders of magnitude smaller than the strain values typically reported in the literature and inferred in this work. Gravitational sag can thus be neglected. Therefore, the reduction in SHG intensity observed near the hole edges is consistent with mechanical stress associated with the clamping boundaries, highlighting the role of the local mechanical state in determining the nonlinear optical response of MoS$_2$.

\begin{figure}
  \centering
  \includegraphics{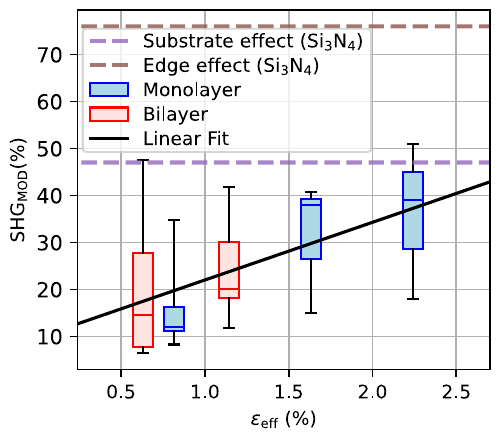}
  \caption{SHG modulation induced in suspended regions of MoS$_2$ monolayers and bilayers as a function of the estimated effective strain, presented as boxplots. Data for each thickness are divided into three and two effective-strain subsets for clarity. In each boxplot, the box spans the interquartile range, the internal line indicates the median, and the whiskers extend to include 100\% of the data. The solid black line represents a linear fit to the data, yielding a modulation sensitivity of $\sim$12\% per 1\% effective strain. Dashed lines indicate the average SHG modulation associated with substrate coupling (violet) and with the clamping boundary effect (brown) observed in the SHG maps.}
  \label{fig:5}
\end{figure}

Analysis of the signal intensity in the supported region, normalized to the maximum observed in the suspended regions, also showed no significant differences across thicknesses. This signal corresponds to SHG$_\textrm{MOD}$ values of $(50 \pm 8)\%$ (monolayer), $(48 \pm 9)\%$ (bilayer), $(45 \pm 8)\%$ (trilayer), and $(44 \pm 9)\%$ (quadrilayer). For comparison with the AFM-induced modulation in Fig.~\ref{fig:5}, the substrate- and edge-related modulations are expressed using the higher-SHG state as the reference, so that $\textrm{SHG}_{\textrm{MOD}}$ consistently represents a decrease in signal. The resulting numerical value therefore depends on the chosen reference intensity.

We next investigate the effect of controlled, localized mechanical deformation on the SHG response of suspended MoS$_2$ monolayers and bilayers using an AFM tip. The protocol used to precisely align the tip with the suspended MoS$_2$ region consists of simultaneously acquiring AFM images and the reflection of the excitation laser from a structural feature such as the holes, then positioning the AFM tip so that both signals coincide spatially. For each measurement, SHG maps were acquired before initiating the contact protocol and after applying pressure to the membrane, in order to confirm that the process does not produce irreversible damage within the force range employed.

\begin{figure}
  \centering
  \includegraphics{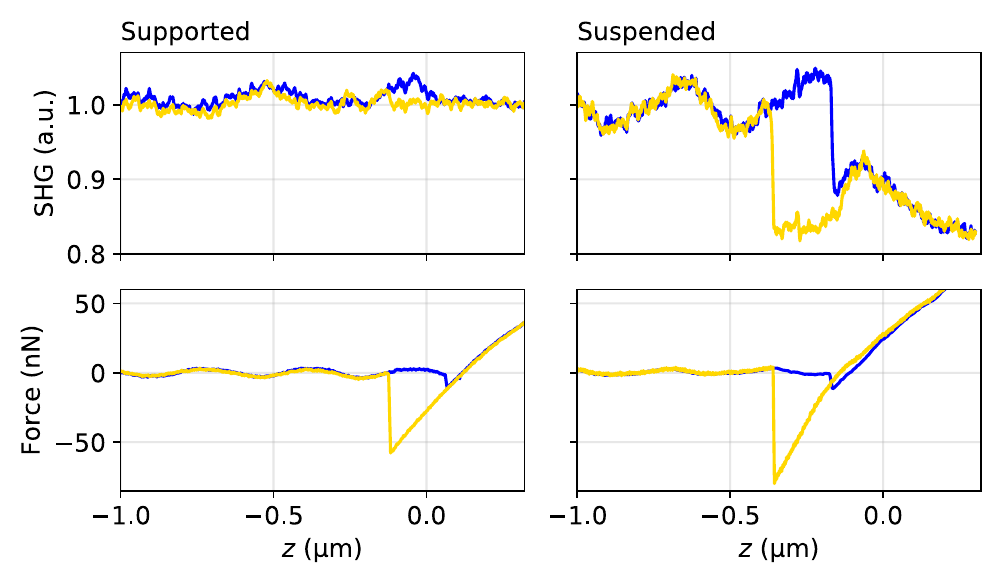}
  \caption{Approach (blue) and retraction (yellow) cycle measurements on supported (left panels) and suspended (right panels) regions of a MoS$_2$ monolayer. Upper panels: SHG intensity normalized to a reference curve acquired with the tip retracted $\sim$50~\textmu{}m from the sample. Lower panels: force exerted on the sample by the AFM tip, derived from the normal deflection signal. Prior to contact, the SHG signal exhibits oscillations arising from interference between the emission collected directly and that reflected off the cantilever. Contact with the suspended membrane modifies the SHG response, while during retraction adhesion between the tip and the MoS$_2$ maintains mechanical deformation until tip detachment.}
  \label{fig:4}
\end{figure}

The signals recorded during the approach--retraction cycle are presented in Fig.~\ref{fig:4}. The figure shows the SHG signal (upper panels) and the force derived from the normal deflection signal (lower panels) during approach (blue) and retraction (yellow) of the stage relative to the tip, determined by the $z$ coordinate. The left panels correspond to a control experiment in which the tip exerts force on MoS$_2$ fully supported on the substrate, while the right panels show representative results obtained with the AFM tip acting at the center of a suspended monolayer region. As the stage moves along $z$, the sample enters and exits the focal plane, producing a relative variation in the detected signal. To account for this variation, the SHG signals are normalized using reference curves obtained using the same $z$-scan protocol with the AFM tip positioned $\sim$50~\textmu{}m away from the sample (details in section~S3 of the Supporting Information).

In the upper panels of Fig.~\ref{fig:4}, oscillations in the SHG signal around the normalization value are observed prior to contact, arising from interference between the directly collected SHG emission and that reflected by the cantilever. These fluctuations, on the order of 5--10\% of the signal, establish a lower bound for identifying mechanically induced modulations. Importantly, the SHG response during approach is not identical across suspended membranes. Once contact is established, increasing indentation modifies the local and extended strain distribution, but the sign and magnitude of the resulting SHG modulation remain sample dependent. Representative approach--retraction cycles illustrating this variability, including cases in which the SHG signal does not decrease upon increasing positive force, are shown in section S3 of the Supporting Information. This behavior may reflect differences in the initial mechanical state of individual membranes, including residual pre-tension, slack or curvature, as well as variations in the local tip--membrane contact geometry. Consequently, a given positive applied force does not necessarily generate the same in-plane strain distribution in different suspended regions.

In contrast, a considerably more reproducible response is observed during retraction. As in conventional force curves, adhesion maintains the tip in contact with the MoS$_2$ while the stage is withdrawn, resulting in a negative cantilever force. In the suspended regions, this adhesive interaction pulls the membrane out of plane and generates tensile in-plane deformation that extends beyond the immediate tip--sample contact. This regime is consistently accompanied by a reduction in SHG intensity, which persists until tip detachment. Although tip–sample adhesion occurs in both supported and suspended MoS$_2$, a significant SHG modulation is observed only in the suspended regions, suggesting that the effect arises from membrane deformation rather than from tip–sample contact alone. After detachment, the SHG signal returns to the level corresponding to the mechanically unperturbed state, demonstrating that the effect is reversible within the explored force range.

To compare the mechanically induced SHG modulation among different suspended regions, we estimate a phenomenological effective strain associated with the AFM-induced deformation. For each approach--retraction cycle, the minimum SHG intensity occurs at the same $z$ position as the minimum force immediately preceding tip detachment. The SHG modulation and the corresponding pull-off force, $F_{\mathrm{pull-off}}$, are therefore determined simultaneously at this point. The magnitude of $F_{\mathrm{pull-off}}$ is used as a characteristic force of the adhesive tip--MoS$_2$ interaction. Although the force established immediately after snap-in is not necessarily identical to the pull-off force because of contact hysteresis, $F_{\mathrm{pull-off}}$ provides a directly measured and reproducible estimate of the characteristic adhesive force sustaining the tip--membrane contact. 
Under these conditions, we use the large-load spherical-indentation result to estimate a characteristic strain scale at the center of the membrane, taking the magnitude of the pull-off force as the characteristic load during retraction.\cite{BHATIA1968} Because the spatial strain distribution during adhesion-mediated retraction is not directly measured, we introduce a simple phenomenological axisymmetric strain profile that decreases smoothly from its estimated central value toward the clamped boundary. The area-weighted average of this profile defines the effective strain:

\begin{equation}
    \varepsilon_{\text{eff}}=
    \frac{1}{3}\left( \frac{|F_{\mathrm{pull-off}}|}{4 \pi E^{\text{2D}} R_\text{tip}}\right)^{1/2}, 
    \label{strain}
\end{equation}

\noindent where $E^{\text{2D}}$ is the two-dimensional Young's modulus: $(180\pm60)$~N/m and $(260\pm70)$~N/m for monolayer and bilayer MoS$_2$, respectively\cite{Bertolazzi2011}, and $R_{\mathrm{tip}}=20$~nm is the nominal AFM tip radius specified by the manufacturer. Under these experimental conditions, $|F_{\mathrm{pull-off}}|$ was typically on the order of 10$^2$ nN. Details of the assumed radial strain profile and spatial averaging are presented in Section~S4 of the Supporting Information. This estimate provides a common strain scale for comparing measurements performed with the same tip geometry and should be interpreted as a phenomenological effective strain rather than as a reconstruction of the complete spatially varying in-plane strain field of the suspended membrane.

Fig.~\ref{fig:5} summarizes the SHG modulation (Eq.~\ref{ec_shg}) as a function of the estimated effective strain (Eq.~\ref{strain}). A clear increase in SHG modulation is observed with increasing $\varepsilon_{\mathrm{eff}}$, showing that stronger adhesion-mediated mechanical deformation is associated with a larger reduction in SHG intensity. The monolayer and non-centrosymmetric bilayer responses show comparable behavior over the explored effective-strain range of approximately 0.5\% to 2.5\%. The dispersion in $\varepsilon_{\mathrm{eff}}$ reflects variations in the measured pull-off force and therefore in the strength of the adhesive tip--sample interaction among different suspended regions. Eleven monolayers and ten bilayers were measured. For clarity, the monolayer and bilayer data were each divided into three and two effective-strain subsets and presented as boxplots using the average $\varepsilon_{\mathrm{eff}}$ of each subset. The box represents the interquartile range, the line within the box represents the median, and the whiskers extend to include 100\% of the data. A linear fit to all measured data yields a characteristic sensitivity of $\sim$12\% SHG modulation per 1\% effective strain.

For comparison, Fig.~\ref{fig:5} also indicates the average SHG modulation associated with substrate coupling and with the clamping boundary effect observed in the SHG maps. These values are included only as references for the magnitude of the optical modulation. Because the strain distributions associated with AFM pulling, substrate coupling, and clamping at the hole boundary are mechanically distinct, the AFM-derived relationship between SHG modulation and effective strain is not used to assign strain values to the substrate- or edge-induced modulation.

In conclusion, we have demonstrated reversible modulation of second-harmonic generation in suspended MoS$_2$ membranes through localized mechanical deformation applied by an AFM tip. While the SHG response during loading depends on the initial mechanical state of individual membranes, adhesion-mediated pulling during retraction consistently produces a reduction in SHG intensity that persists until tip detachment. By relating the measured pull-off force to a phenomenological estimate of effective strain, we observe a systematic increase in SHG modulation with increasing mechanical perturbation in both monolayer and bilayer MoS$_2$. Furthermore, suspended regions exhibit enhanced SHG intensity relative to substrate-supported regions, with pronounced suppression near the clamping boundaries, consistent with the influence of local mechanical conditions on the nonlinear optical response. Together, these findings establish localized mechanical deformation, particularly adhesion-mediated pulling, as a means of reversibly controlling second-harmonic generation in atomically thin membranes.



\section*{Funding Sources}
This work was partially supported by ANPCyT (PICT 2021 IA-363), CONICET (PIP 112-20200101465), UBACyT (Proyecto 20020220200078BA), and Red Federal de Alto Impacto (FFFLASH; CONVE-2023-10189190). A.V.B. acknowledges funding support from the Alexander von Humboldt Foundation through the 2023 Georg Forster Award.

\section*{Acknowledgements}

The authors thank Dr. I. Abdelwahab (Harvard University) for providing the bulk crystals used for exfoliation, Mr. Gabriel Bosch (Centro Atómico Constituyentes, CNEA) for assistance with fabrication of the microstructured substrate, and Camila Sanchez for assistance in the transfer process.

\section*{Supporting information}

This material is available free of charge via the Internet at http://pubs.acs.org
\begin{itemize}
    \item Additional details on sample fabrication, transfer, and characterization; a comparison between SHG and PL measurements on microstructured substrates; normalization procedure for the SHG curves presented in the main text; representative approach–retraction cycles illustrating the variability in the SHG modulation response; and details on the calculation of the effective strain.
\end{itemize}

\bibliography{Bibliography.bib}

\end{document}